\documentclass[12pt,preprint]{aastex}

\usepackage{emulateapj5}
\usepackage{epsf}

\newcommand{\tx}[1]{\textrm{#1}}

\newcommand{\kms}{km~$\tx{s}^{-1}$}

\newcommand{\mlsun}{M$_{\odot}$/L$_{\odot,V}$}

\newenvironment{inlinefigure}{
\def\@captype{figure}
\noindent\begin{minipage}{0.999\linewidth}\begin{center}}
{\end{center}\end{minipage}\smallskip}

\newenvironment{inlinetable}{
\def\@captype{table}
\noindent\begin{minipage}{0.999\linewidth}\begin{center}}
{\end{center}\end{minipage}\smallskip}

\slugcomment{ApJ Letters, in press}

\shorttitle{The dark matter profile of MS2137-23}
\shortauthors{Sand, Treu, \& Ellis}

\begin{document}

\title{The dark matter density profile of the lensing cluster
MS2137-23: a test of the cold dark matter paradigm$^1$}
\footnotetext[1]{Based on observations collected at the Keck
Observatory, which is operated jointly by the California Institute of
Technology and the University of California, and with the NASA/ESA
HST, obtained at STScI, which is operated by AURA, under NASA contract
NAS5-26555.}

\author{David J. Sand}
\affil{California Institute of Technology,
Physics, mailcode 103--33, Pasadena, CA 91125}
\author{Tommaso Treu \& Richard S. Ellis}
\affil{California Institute of Technology,
Astronomy, mailcode 105--24, Pasadena, CA 91125}

\begin{abstract}
We present new spectroscopic observations of the gravitational arcs
and the brightest cluster galaxy (BCG) in the cluster MS2137-23
($z=0.313$) obtained with the Echelle Spectrograph and Imager on the
Keck II telescope. We find that the tangential and radial arcs arise
from sources at almost identical redshifts ($z=1.501,1.502$). We
combine the measured stellar velocity dispersion profile of the BCG
with a lensing analysis to constrain the distribution of dark and
stellar matter in the central 100 kpc of the cluster. Our data
indicate a remarkably flat inner slope for the dark matter profile,
$\rho_d\propto r^{-\beta}$, with $\beta<0.9$ at 99\% CL. Steep inner
slopes obtained in cold dark matter cosmological simulations -- such
as Navarro Frenk \& White ($\beta=1$) or Moore (1.5) universal dark
matter profiles -- are ruled out at better than $99$\%CL. As baryon
collapse is likely to have steepened the dark matter profile from its
original form, our data provides a powerful test of the cold dark
matter paradigm at the cluster mass scale.
\end{abstract}


\section{Introduction}

A fundamental result arising from cold dark matter (CDM) numerical
simulations is that the density profiles of DM halos are
universal in form across a wide range of mass scales from dwarf
galaxies to clusters of galaxies (Navarro, Frenk \& White 1997,
hereafter NFW). Internal to some scale radius $r_{sc}$, the dark
matter profile assumes a power law form, $\rho_{d} \propto
r^{-\beta}$. Whilst there is some dispute amongst the simulators about
the precise value of $\beta$ with values ranging from 1.0 to 1.5,
(Moore et al.\ 1998, hereafter M98; Ghigna et al.\ 2000, Power et
al. 2002), a clear measurement of $\beta$ in a range of objects would
offer a powerful test of the CDM paradigm.

The largest observational effort in this respect to date has been via
dynamical studies of low surface brightness (LSB) and dwarf galaxies,
suggesting softer ($\beta<1$) DM cores than expected on the basis of
the numerical simulations (e.g. de Blok \& Bosma 2002, Salucci \&
Burkert 2000), although the issue remains somewhat controversial
(e.g. van den Bosch \& Swaters 2001). Similar tests have recently been
extended to regular spiral (Jimenez, Verde \& Oh 2002) and elliptical
galaxies (Treu \& Koopmans 2002).  Some observational constraints are
available at the scale of massive clusters, from lensing (e.g. Tyson,
Kochanski \& Dell'Antonio 1998; Williams et al.\ 1999; Smith et al.\
2001), X-ray analysis (Mahdavi \& Geller 2001) and dynamics of cD
galaxies (Kelson et al. 2002).  Since massive clusters probe a totally
different scale and physical conditions than galaxies, it is crucial
to understand their mass distribution to test the universality of the
DM profiles.

In this paper we present the first application of a new method to
determine the luminous and dark mass distribution in the inner regions
of massive clusters with giant arcs around a central BCG. The method
combines lensing analysis with stellar kinematical measurements of the
BCG. The two ingredients provide complementary information on the
relevant scales ($\sim 100$ kpc), allowing us to disentangle the
luminous and dark components of the total mass distribution.

We have chosen the cluster MS2137-23 as a first application of our
method since it is an approximately round system, has an isolated BCG
and a very well-studied arc system. Fort et al.\ (1992) first pointed
out the potential significance of the radial and tangential
gravitational arcs as a means of constraining the mass distribution on
$\simeq$100 kpc scales, and mass models have been developed
subsequently (Mellier et al.\ 1993, Hammer et al.\ 1997; hereafter
M93, H97). The redshifts of the radial and tangential arcs were
predicted to lie in the range 1$<z<$2 (M93). A key issue in the
earlier work is whether the radial arc is, in fact, a lensed
feature. M93 and Miralde-Escud\'e (1995; hereafter ME95) also point
out the importance of determining the stellar velocity dispersion
profile of the BCG to weigh the stellar contribution to the mass.

Following the earlier suggestions, we present new observations of the
cluster MS2137-23 with the Keck II telescope.  We provide
spectroscopic confirmation of the arcs and measure a velocity
dispersion profile for the central BCG.  The spectroscopic data are
used together with archival HST images to constrain the luminous and
DM distribution of the cluster. In the following, $r$ is the radial
coordinate in 3-D space, while $R$ is the radial coordinate in 2-D
projected space. We adopt H$_0$=65~km s$^{-1}$,Mpc$^{-1}$,
$\Omega_m=0.3$ and $\Omega_{\Lambda}=0.7$ for the cosmological
parameters.

\clearpage
\section{Observations}
\subsection{Keck Spectroscopy}

We observed MS2137-23 using the Echelle Spectrograph and Imager (ESI;
Sheinis et al.\ 2002) on the W.~M. Keck-II telescope for a total
integration time of 4900s (2$\times$1800s + 1300s) on 21 July,
2001. The seeing was $0\farcs6$ and the $1\farcs25\times20''$ slit was
oriented North-South to include the BCG, radial arc, and tangential
arc (Figure 1). The spectroscopic goals were two-fold: a determination
of the redshift of the arcs and a measurement of the internal
kinematics of the central galaxy. An {\sc iraf} package was developed
for the specific task of removing echelle distortions while preserving
the 2-D shape of the spectrum essential for the latter goal (EASI2D,
Sand et al.\ 2002, in preparation).  The instrumental resolution of
ESI was measured from unblended sky lines to be 30$\pm$7 km s$^{-1}$.

The velocity dispersion profile of the BCG (Figure 2) was measured
using spectral templates based on several G-K giants observed with a
$0\farcs3$ slit. These were smoothed to match the instrumental
resolution of the $1\farcs25$ slit and redshifted to that of the BCG
($z$=0.313). Analysis was restricted to a region around the G band by
virtue of the high signal/noise and minimal effect of sky line
residuals, using the Gauss-Hermite pixel-fitting software (van der
Marel 1994). The error bars shown in Fig. 2 represent a combination of
uncertainties arising from Poisson noise and systematics, the latter
determined from the scatter observed using different templates and
continuum fits.

\begin{inlinefigure}
\begin{center}
\resizebox{\textwidth}{!}{\includegraphics{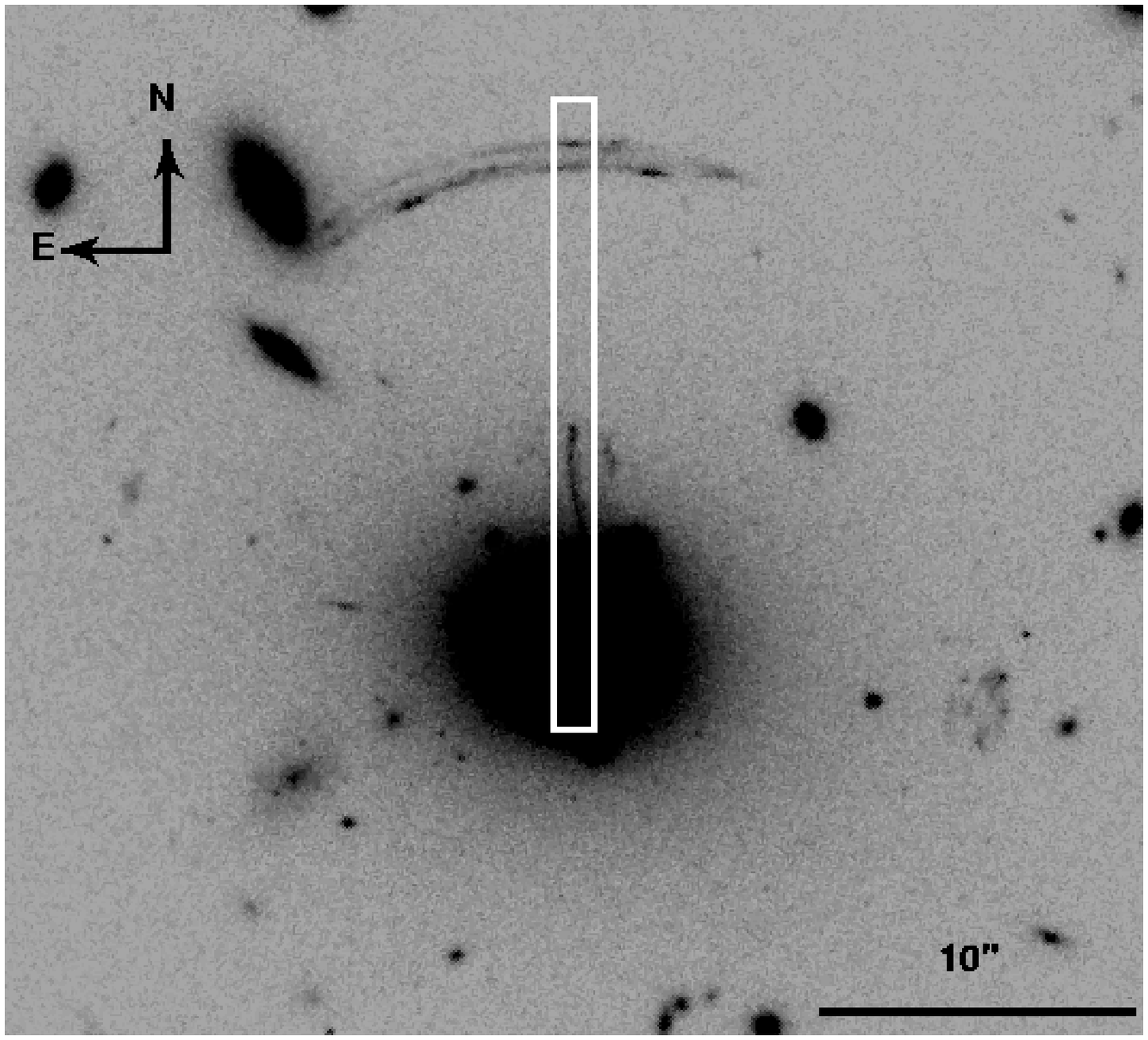}}
\end{center}
\figcaption{HST F702W image of MS2137-23. The rectangular box shows
the dimensions ($1\farcs25\times20''$) and position of the ESI slit
used to make the observations. The BCG, the radial and tangential arc
are clearly visible at the bottom, center, and upper end of the slit.
\label{fig:ms2137pic}}
\end{inlinefigure}

The high spectral resolution of ESI proved crucial in clinching the
redshifts of the arcs as the emission lines are located in a crowded
region of OH sky background. The two top panels in
Fig.~\ref{fig:vdprof} show the relevant portion of the ESI spectra for
the tangential and radial arcs, with the observed emission lines
identified as the [OII] doublet at $z=1.501$ and $z=1.502$
respectively.  The [OII] doublet is clearly resolved for the
tangential arc and it is reasonable to suppose that the missing
component for the radial arc is obscured by sky emission. No other
lines are detected on either spectra, down to the blue cutoff of ESI
at $\sim3900 $\AA.  This makes it unlikely that the single line
observed for the radial arc is any of the common lines such as
H$\alpha$, H$\beta$, [OIII]4959, 5007, CIV1549, HeII1640, C[III]1909
because bluer lines would be detected assuming typical flux
ratios. The identification of the line with Ly$\alpha$ at $z=6.66$ is
also unlikely given that the arc is detected in the HST F702W image
(see below).  Detailed modeling of MS2137-23 based on the image
configurations predicted that the sources for the arcs would be at
nearly the same redshift (M93, H97).
\begin{inlinefigure}
\begin{center}
\resizebox{\textwidth}{!}{\includegraphics{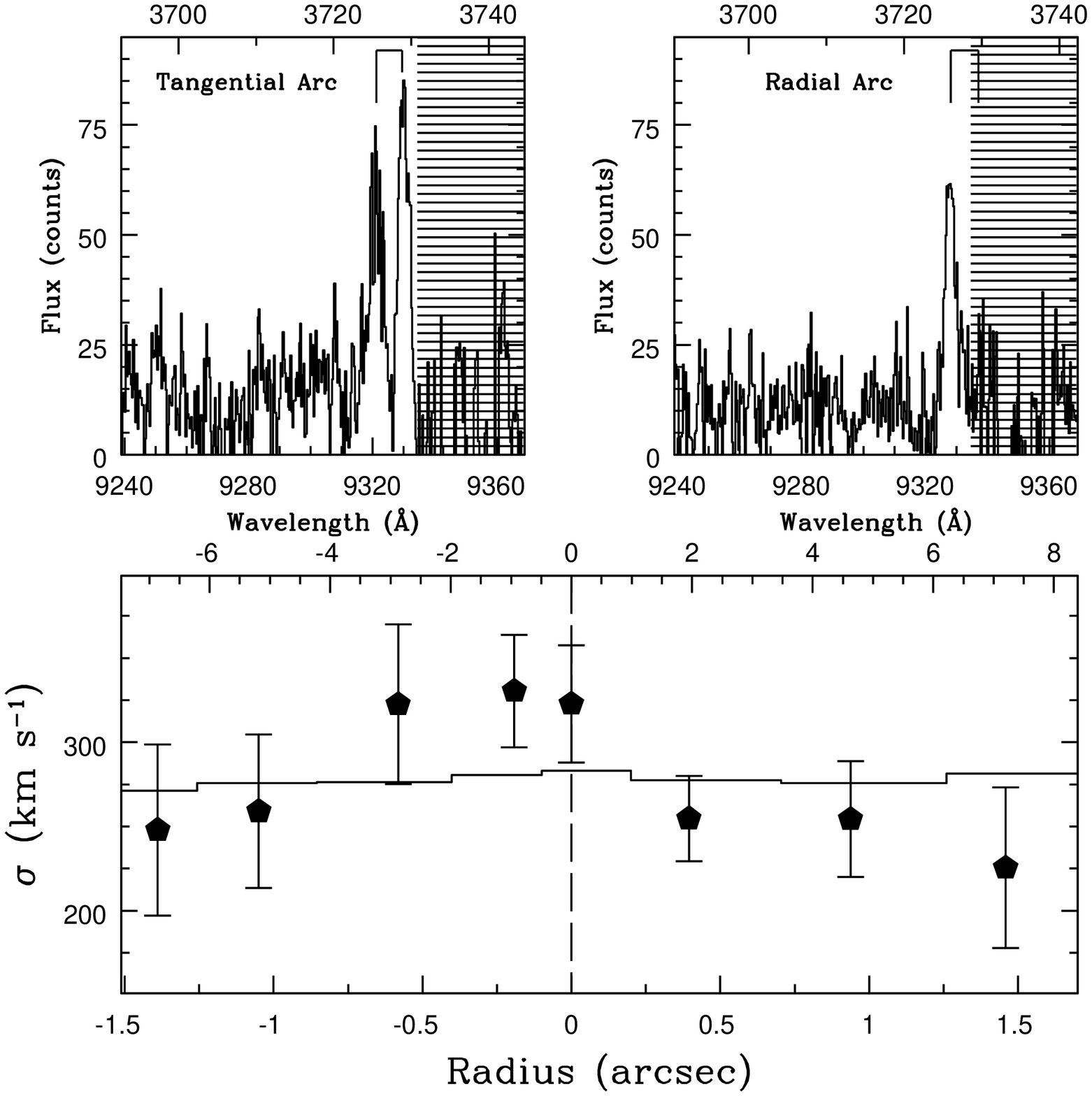}}
\end{center}
\figcaption{Spectroscopic results: (Top) Strong emission lines
detected in the spectra of the tangential and radial arcs. These
are identified as [OII]3726,3729 at $z=1.501$ and $z=1.502$
respectively (marked). It is argued that the missing 3729$\AA$ 
line in the radial arc is obscured by sky emission. (Bottom)
Stellar velocity dispersion profile of the brightest cluster
galaxy (points with error bars). The superimposed histogram shows
the profile of the best fitting Jaffe + generalized NFW mass model
(see Sec.~3 for details), taking into account the effects of
seeing ($0\farcs6$), slit width, and radial binning.
\label{fig:vdprof}}
\end{inlinefigure}
\subsection{Hubble Space Telescope Imaging}

HST WFPC2 images of MS2137-23 (GO 5402, PI: Gioia), comprising 10
F702W exposures with a total integration time of 22.2ks, were used to
measure the surface photometry of the BCG and to locate arc positions.
The exposures were reduced in a standard way, using the {\sc iraf}
package {\sc drizzle} (Fruchter \& Hook 2002).

The circularized surface brightness profile (in agreement with H97)
was obtained using the {\sc iraf} task {\sc ellipse} and a fit
performed as described in Treu et al.  (1998, 2001a) taking into
account the HST point-spread function. The best fitting $R^{1/4}$
parameters are summarized in Table~1.  To convert from F702W
magnitudes to V magnitudes a k-color correction was calculated using
the same method as Treu et al.  (1999).  Rest frame photometric
quantities were corrected for Galactic extinction using $A_{F702W}$ =
2.435E(B-V)=0.122 (Schlegel et al.\ 1998).

\begin{inlinetable}
\centering
\begin{tabular}{lr}
\hline
\hline
 Redshift (BCG)          & $0.313\pm0.001$ \\
Radial critical line     & $4\farcs5\pm0\farcs3$ \\
Tangential critical line & $15\farcs35\pm0\farcs20$ \\
 (1-b/a)$_e$ & $0.17\pm0.01$\\
 F702W (mag) & $16.48\pm0.07$ \\
 SB$_{e,F702W}$ (mag/arcsec$^2$) & $23.58\pm0.34$ \\
 R$_{e,F702W}$ & $5\farcs02\pm0\farcs50$ \\
 M$_V$ (mag) & $-24.38\pm0.09$ \\
 SB$_{e,V}$ (mag/arcsec$^2$) & $22.76\pm0.34$ \\
 R$_{e,V}$ (kpc) & $24.80\pm1.68$ \\
\hline
\hline
\end{tabular}
\end{inlinetable}
\vspace{0.2cm}

\noindent{\footnotesize TABLE~1: Relevant spectro-photometric
quantities}

\section{Luminous and dark matter distribution}

We now combine the observed spectroscopic and photometric data to
constrain the matter distribution in the central region of
MS2137-23. First we introduce a simple two-component spherical mass
model comprising the stellar mass of the BCG and a DM halo
(Sec.~3.1). We then constrain the free parameters of the model using
the position of the critical lines (3.2) and the velocity dispersion
profile (3.3).

\subsection{Mass model}

For the luminous component we used a Jaffe (1983)
\begin{equation}
\label{eq:jaffe}
\rho_L(r)=\frac{M_{L} r_{J}}{4 \pi r^{2} (r_{J} + r)^{2}},
\end{equation}
mass density profile of total mass M$_L$, which reproduces
well\footnote{As a check of the results, we also considered a
Hernquist (1990) luminous mass distribution. The results on $\beta$
(see below) are virtually unchanged, while slightly larger values of
M/L for the stellar component are obtained.} the observed surface
brightness profile (with $R_{e}$ = 0.76$r_{J}$). The DM halo
is modeled as,
\begin{equation}
\label{eq:gnfw}
\rho_d(r)=\frac{\rho_{c} \delta_{c}}{(r/r_{sc})^{\beta}(1+(r/r_{sc}))^{3-\beta}},
\end{equation}
representing a generalization of the CDM-motivated halos, with an
inner slope $\beta$ (NFW and M98 correspond to $\beta=1,1.5$
respectively; $\rho_c$ is the critical density).  We assume that the
BCG lies at the center of the overall potential.

For a given stellar mass-to-light ratio $M_*$/$L_V$ both $M_{L}$ and
$r_{J}$ can be deduced from the surface photometry leaving 4 free
parameters in our mass model: 1) $M_*$/$L_V$; 2) the inner slope of
the DM profile $\beta$; 3) the DM density scale $\delta_{c}$;
and 4) the DM scale radius $r_{sc}$.

\subsection{Gravitational Lensing}

Given our two-component spherical model, we adopted a simple lensing
analysis using only the positions and redshifts of the radial and
tangential arcs (see Bartelmann 1996).

The locations of the radial and tangential arcs can be estimated by
calculating the position of the corresponding radial and tangential
critical curves of the projected mass distribution. The Jacobian
matrix of the lens mapping has two eigenvalues, $\lambda_r= 1 -
\frac{d}{dx}\frac{m}{x}$ and $\lambda_t= 1 - \frac{m}{x^2}$, where
$x=R/r_{sc}$ and $m$ is a dimensionless function proportional to the
mass inside projected dimensionless radius $x$ (see, e.~g., Bartelmann
1996, Schneider, Ehlers, Falco 1992).  Tangential and radial critical
curves occur when $\lambda_t=0$ and $\lambda_r=0$, respectively.  In
practice, the position of the tangential arc constrains the {\em total
enclosed mass}, while the position of the radial arc constrains its
{\em derivative}. A proper account of ellipticity is essential for
detailed lens modeling where the shape, magnification, and morphology
of multiple lensed images is being reproduced.  ME95, however,
considered several different simple mass models where only the
position of the radial and tangential critical lines were being
measured and found that the position of the two was affected very
little by the introduction of ellipticity. Therefore, we conclude that
a spherical model is appropriate for our analysis.

For every set of free parameters \{$M_*/L_V$, $\beta$, $\delta_c$,
$r_{sc}$ \}, we can compute the predicted position of the arcs, find
the likelihood assuming gaussian distributions, and constrain the
acceptable mass models. The largest radius at which the mass is probed
is that corresponding to the location of the tangential arc (75.8
kpc). Now $r_{sc}$ is expected to be much greater than 100 kpc in CDM
clusters (Bullock et al.\ 2001; see also Wu 2000). In this case, the
location of the critical lines depends only marginally on $r_{sc}$ and
the combined luminous and dark density profile has only 3 free
parameters (we fix $r_{sc}=400$ kpc in the following).  Fig.~3 shows
the likelihood contours (68\%, 95\% and 99\%) in $\beta$
vs. $M_*$/$L_{V}$ space, obtained using the likelihood ratio statistic
after marginalization with respect to $\delta_c$. Note that with
lensing alone we can rule out with greater than 95\% confidence a M98
DM density profile ($\beta$=1.5).

\subsection{Lensing + Dynamics}

The full power of our analysis is only realized when we combine the
earlier constraints with those made by measuring the
spatially-resolved stellar velocity dispersion profile of the BCG.
Given our two-component mass model (Section 3.1), we solved the
spherical Jeans equation (e.g. Binney \& Tremaine 1987) assuming an
isotropic velocity ellipsoid for the luminous
component.\footnote{Spherical dynamical models have been shown to
reproduce accurately the kinematics of slightly elongated galaxies
like the BCG (e.g. Kronawitter et al.\ 2000).}

The assumption of isotropy in the region probed (within 30 \% of
$R_{e}$) we justify both on theoretical and observational
grounds. Numerical simulations (e.~g. van Albada 1982) and
observations (e.g. Gerhard et al.\ 2001; Koopmans \& Treu 2002) appear
to rule out significant tangential anisotropy and permitting some
radial anisotropy only at large radii.  Strong radial anisotropy in
the very central regions can also be ruled out based on the grounds of
stability (e.~g. Merritt \& Aguilar 1985; Stiavelli \& Sparke 1991)
and consistency requirements (Ciotti 1999).  However, as a check, we
ran Osipkov-Merritt (Osipkov 1979; Merritt 1985a,b) models with
anisotropy radius set equal to progressively lower radii.  Moving the
anisotropy radius towards zero pushes the likelihood contours towards
lower values of $\beta$.

For each set of parameters in the lens model, we computed the
likelihood given the velocity dispersion profile, taking into account
the effects of seeing, radial binning, and finite slit width. The
total likelihood was computed by multiplying the velocity dispersion
likelihood by the likelihood obtained with the lensing analysis.  The
bottom panel in Fig.~3 shows the final results of the combined
analysis.  With 99\% confidence, $M_*$/$L_{V}$ lies between 2.3 and
3.7 (broadly consistent with local values after passive evolution),
with the inner DM slope ($\beta$) lying between 0.05 and 0.8, flatter
than that expected from CDM simulations. The best-fitting parameters
are $M_*/L_V=3.1$ \mlsun, $\beta=0.35$, $\delta_c=24000$.

\begin{inlinefigure}
\begin{center}
\resizebox{\textwidth}{!}{\includegraphics{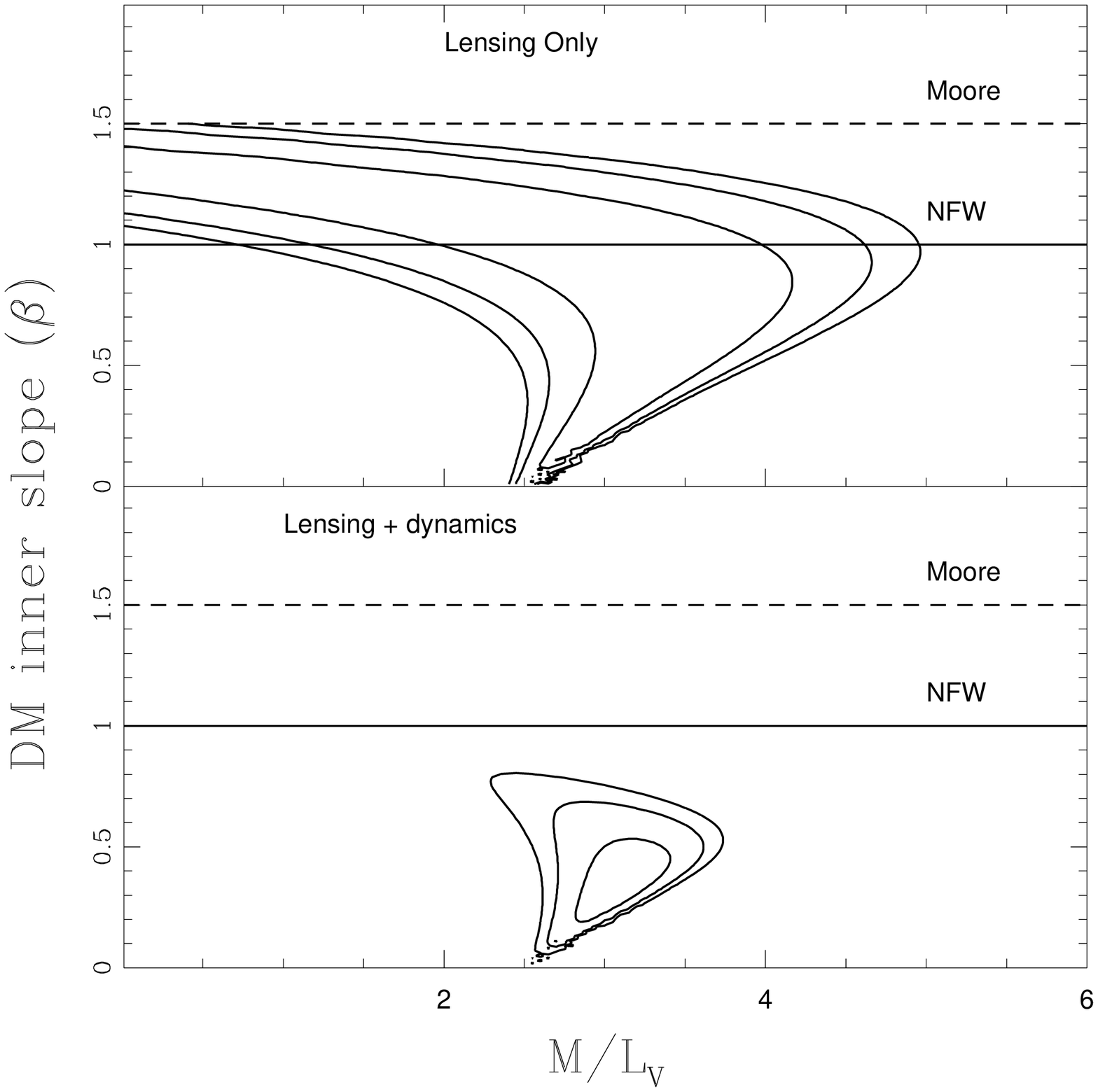}}
\end{center}
\figcaption{Likelihood contours (68\%, 95\%, and 99\%) obtained for
the mass modeling of MS2137-23 with a Jaffe luminous distribution plus
a generalized NFW DM distribution. (Top): Contours obtained from the
position of the radial and tangential arcs alone.  Note that a M98
($\beta=1.5$) profile is excluded at the 95\% level. (Bottom):
Contours obtained including the measured velocity dispersion profile.
Note the improved constraints on the mass parameters and that NFW
profiles are clearly ruled out at the 99\% level.
\label{fig:mlplot}}
\end{inlinefigure}

To check our results, we changed $R_{e}$ by 10\% with a negligible
effect on the likelihood contours. Similarly, the contours are
virtually unchanged by offsets of $0\farcs5$ in the position of the
tangential arc and by changing the seeing by 30\%. Changing the
position of the radial critical line by $\pm0\farcs5$ shifts the
likelihood contours by $\pm$0.1 in the $\beta$ direction.

Systematic offsets of the velocity dispersion profile due to template
mismatch and poor continuum fitting introduce correlation between the
kinematic points that are not considered in the likelihood ratio
analysis.  To investigate this, we have repeated the analysis with the
velocity dispersion profile shifted by the estimated systematic error
($\sim 15$ \kms). A lower overall velocity dispersion profiles shifts
our likelihood contours towards lower $\beta$ ($\beta<0.65$ 99 \% CL),
and viceversa ($\beta<0.9$ 99 \%CL): our hard 99\% CL upper limit is
$\beta<0.9$.

\section{Summary and Discussion}

We have secured spectroscopic redshifts for the radial and tangential
arcs in the cluster MS 2137-23 and determined the velocity dispersion
profile of the BCG to a physical radius of $\simeq$8 kpc. We have
combined these measurements with the lensing geometry in order to
construct a self-consistent model of the mass distribution in the
cluster core. Using a spherically symmetric luminous and DM
mass distribution, we rule out the presence of a DM halo with an inner
slope $\beta>0.9$ at greater than 99\% confidence, including
systematics.  M98 and NFW-type halos with $\beta\ge1.0$ are
inconsistent with the mass distribution in the core of MS2137-23.

Since the infall of baryons associated with the BCG are likely to {\em
steepen} the DM halo (Blumenthal et al 1986; Mo, Mao and White 1998),
our measured profile may imply the original DM profile was even
flatter.  A full modeling of this process is beyond the scope of this
letter, but this strengthens our conclusion that the inner regions of
the DM halo of MS2137-23 cannot be described by CDM-motivated
universal halos. A potential concern is that our models have only two
mass components, but X-ray emitting gas could be a non-negligible
third massive component.  Using ROSAT observations of MS2137-23
(Ettori \& Fabian 1999) we estimate that removing the X-ray component
will steepen the resulting DM halo slope by less than $\simeq0.1$ and
therefore does not change dramatically the result.

Finally, individual halo shapes can depart from the ensemble average
behavior. Therefore it is necessary to apply such a test to a sample
of clusters. Our simple method is applicable to all approximately
round clusters with a massive galaxy at their center, provided that
they have at least a giant tangential arc (radial arcs further enhance
the sensitivity but are not required). We are in the process of
collecting data for a dozen clusters with the aim of performing such a
statistical test.

\acknowledgements

We are grateful to A.~Benson, J.-P. Kneib, L.~Koopmans for insightful
discussions and comments on this project, and to P.~Shopbell for
numerous computer tips. We acknowledge the use of the Gauss-Hermite
Pixel Fitting Software developed by R.~P.~van der Marel.  We
acknowledge financial support for proposal number HST-AR-09527
provided by NASA through a grant from STScI, which is operated by
AURA, under NASA contract NAS5-26555.  Finally, the authors wish to
recognize and acknowledge the cultural role and reverence that the
summit of Mauna Kea has always had within the indigenous Hawaiian
community.  We are most fortunate to have the opportunity to conduct
observations from this mountain.

\end{document}